\documentstyle[preprint]{ptptex}

\newcommand{\bl}[1]{\mbox{\boldmath{$ #1$}}}
\newcommand{\blsmall}[1]{\mbox{\boldmath{ \footnotesize $ #1$}}}
\newcommand{\epsl}{\varepsilon}
\newcommand{\f}[2]{\frac{#1}{#2}}
\newcommand{\dd}{\partial}

\preprintnumber[3cm]{KNK-01I23}

\title{
Statistics of Composite Systems \\ and \\ Anyons 
in the Fractional Quantum Hall Effect
}

\author{
Hitoshi {\sc Ito}\footnote{
 E-mail: itoh@phys.kindai.ac.jp}
}

\inst{
Department of Physics, Faculty of Science and
 Engineering, \\ Kinki University, Higashi-Osaka, 577-8502, Japan \\
 December, 2001}

\recdate{
}

\abst{
The commutation relations of composite fields are studied in the 3, 2 and 1 spatial dimensions. It is shown that the field of an atom consisting of a nucleus field and an electron field satisfies, in the space-like asymptotic limit, the canonical commutation relations within the sub-Fock space of the atom. The field-particle duality in the bound state is discussed from a statistical point of view.  Then, the commutation relations of the scalar object in the Schwinger (Thirring) model are briefly discussed and they are shown to be consistent with its interpretation as the Nambu-Goldstone boson. 

The composite anyon fields are shown to satisfy the proper anyonic commutation relations with additive phase exponents. Then, quasiparticle picture of the anyons is clarified under the restriction of this additivity. 
The difference between field and particle characteristics becomes more prominent in the 2-dimensional space. 
It is argued that the hierarchy of the fractional quantum Hall effect can be rather simply understood by utilizing the quasiparticle characters of the anyons in the case that the background-boson gauge is assumed. To contrast with this case, the coposite fermion theories are critically reviewed.
}

\begin{document}

\maketitle

\section{Introduction}

More than forty years ago, composite systems in quantum field theory were investigated. 
In such a study, one starts from a local scalar field, say $A(x)$, and assumes the existence of discrete eigenvalues $m^2$ and $M^2$ of the 4-momentum squared, $P^2$, where $m$ is the mass of the original field $A$.
If $\langle A(x)A(y)\mid P^2=M^2\rangle \neq 0$ there may be a composite (bound)  state of mass $M$. Then, one defines a bilocal field
\begin{equation}
  B(x,\epsl)=TA(x+\epsl)A(x-\epsl)        \label{a1}
\end{equation}
representing this state, where $T$ denotes the time-ordered product.
Zimmermann showed, with some mathematical assumption, that the asymptotic ($t\to\pm\infty$) field of $B$ satisfies the proper commutation relation in the limit $\epsl\to 0$, if it is suitablly normalized.\cite{Zimmermann58}
 One can infer from this construction that composite systems can be described  by the local field operators, and there are no differences between elementary and composite particles in constructing the $S$ matrix elements.
The idea of boot strapping (nuclear democracy) emerged from this observation.

However, the successes of gauge theories have changed drastically the framework of elementary particle theory. 
As a result, the quantum fields revived, and the hierarchy structure of the gauge interactions was gradually recognized. 
Field theory in its present form is an effective theory for each level of class in the hierarchy.

There is another hierarchy of compositeness in nature,\cite{Sakata61} which has been revealed through the success of the composite models of elementary particles. The hierarchy here consists of the classes of quarks, hadrons, neuclei, atoms and so on.
 The level of a class is specified by the energy scale, indicating the limit of applicability of the theory governing it.
We, further, believe that the theories describing these classes are quantum field theories and there exist elementary fields for each class, which are constructed from the elementary fields of a deeper class.
The most instructive example of this interpolating mechanism may be provided by considering the class of an atom. 
An atom is a composite system that consists of a nucleus field and an electron field interacting through a photon field. Then, we construct the atom field as a composite field of elementary fields of a deeper class. In this respect, we are especially interested in the statistical properties of the composite system, since we feel it sometimes difficult to understand the statistical properties in the framework of particle quantum mechanics.
 For example, the nucleus of the hydrogen atom is a fermion and it changes into a boson by acquiring another fermion. This is mysterious from the viewpoint of the particle quantum mechanics.\footnote{
A gas of the H atoms undergoes Bose-Einstein condensation. This means that the protons become distributed like bosons, since their positions almost coincide with those of the atoms. 
}

The difference between quantum field theory and many-particle quantum mechanics is more prominent in the world of the 2 spatial dimensions. Its topological structure allows the exotic statistics of the field operators, which cannot always be realized by a (quasi)particle having definite values of the flux and the charge. Only if we can define the attributes of such a composite consistently, we can call it a quasi-particle anyon. The first purpose of the present paper is to give unified consideration to the essential points of the field-particle duality in composite systems. We also emphasize the topological difference between 3 and 2 spatial dimensions. The main issue is to determine how to understand the Chern-Simons (CS) gauge field that governs the statistics of the 2-dimensional space. We consider it to be the field of a boundary condition to be eliminated in the final physical results.

The complicated nature of the statistics of anyons sometimes confuses theoretical understanding of the related phenomena. An example is the fractional quantum Hall effect (FQHE). Our second purpose is to clarify the hierarchical structure of the fractional quantum Hall state (FQHS) from the point of view of the CS gauge theory of anyons, which has not always been recognized correctly.

We study the field aspects of the hydrogen atom in \S \ref{atom} and derive the canonical commutation relations for the atom field. Section \ref{any} is devoted to studying the anyon fields in 2 spatial dimensions. We show the additivity of the phase exponents in the commutation relations, which restricts possible charges and fluxes of quasi-particles. Some aspects of anyons in the FQHE are sketched on this basis. In 1 spatial dimension, a composite field has commutation relations with rather clear-cut structure, which is shown briefly in \S \ref{one}. In \S \ref{hier} we study the hierarchy structure of the FQHS, placing emphasis on the special nature of the CS gauge field. The composite fermion theory is critically reviewed from this standpoint. Finally, we make some conceptual remarks on the subject in the last section.

\section{The atom field  \label{atom}}

The symmetries of the many-body wave function were investigated by Ehrenfest and Oppenheimer\cite{Ehrenfest30} using interesting considerations without referring to field quantization. We now study the same subject in the framework of quantum field theory, intending to make clear the field aspects of these symmetries. We consider here the simplest case of the hydrogen atom.

We first introduce the composite field $\Psi$ through the equation
\begin{equation}
   \Psi(x_1,x_2)=T\psi(x_1)\phi(x_2),       \label{a2}
\end{equation}
where $\psi(x_1)$  and $\phi(x_2)$ are the elementary fields of the nucleus and the electron respectively in the Heisenberg picture. We then define the Bethe-Salpeter amplitude $\langle 0|\Psi(x_1,x_2)|2\rangle$ for the two-particle states and obtain the state of the H-atom by solving the BS equation for it. The nonrelativistic approximation suffices for the present purpose, and we are interested in only the wave function of the ground state, whose Fourier transform is denoted by $g_{mm'}(\bl{k}_1,\bl{k}_2)$, where $m$ and $m'$ are the indices of the spin of the nucleus and the electron, respectively. We note, however, that these indices are dummies since the spin of the nucleus is frozen in the nonrelativistic description, and therefore the spin degrees freedom of the electron can be neglected in the ground $S$ state. 

We next reconstruct the composite field operator by including the bound-state amplitude. If the total and the relative momenta are $K=(K_0,\bl{K})$ and $k=(k_0,\bl{k})$ respectively, the contribution of the bound state to the annihilation part is given by
\begin{eqnarray}
 \lefteqn{ \Psi(X,x) = Ce^{iKX}\sum_{mm'}\int d^3k}  \nonumber \\
   & & \times g_{mm'}(\bl{K},\bl{k})
    a_m(\eta_1\bl{K}-\bl{k})b_{m'}(\eta_2\bl{K}+\bl{k})
                            \exp{(i\bl{k}\cdot\bl{x})},
                     \quad \eta_1+\eta_2=1,     \label{a3} 
\end{eqnarray}
where $C$ is a normalization factor and $a_m(\bl{k}_1)$ and $b_{m'}(\bl{k}_2)$ are the annihilation operators for the nucleus and the electron respectively, which satisfy the anti-commutation relations
\begin{equation}
    \{a_m(\bl{k}_1),a^\dagger_{m'}(\bl{k}_2)\}=
      \delta_{mm'}\delta(\bl{k}_1-\bl{k}_2), \quad \mbox{etc.}   \label{a4}
\end{equation}
Now, when we observe a stable atom from some great distance, we can ignore the scale of the relative coordinate $\bl{x}$, and the atom is represented by the wave function at the origin.
\footnote{
We should use a relativistic equation in deeper classes of the compositeness hierarchy. Then, the wave function at the origin becomes a divergent quantity in some cases, and we have to renormalize it.\cite{Ito82}
}
 Following Haag, we refer to the ignoring of the relative coordinate as the ``space-like asymptotic limit''.\cite{Haag58}

The bound state is represented by a local field operator in the space-like asymptotic limit. Its annihilation part is given by
\begin{equation}
  \Psi(X,0)=CA(\bl{K})e^{iKX},  \quad 
   A(\bl{K})=\int g(\bl{K},\bl{k})a(\eta_1\bl{K}-\bl{k})b(\eta_2\bl{K}+\bl{k})d^3k,  
                               \label{a5}
\end{equation}
where the dummy spin indices are omitted. $A(\bl{K})$ satisfies commutation relations
\begin{equation}
   [A(\bl{K}),A(\bl{K}')]=[A^\dagger(\bl{K}),A^\dagger(\bl{K}')]=0  \label{cc1}
\end{equation}
and
\begin{eqnarray}
 \lefteqn{ [A(\bl{K}),A^\dagger(\bl{K}')] = \delta(\bl{K}-\bl{K}')} 
      \nonumber \\ & & 
   -\int d^3k g(\bl{K},\bl{k})g^*(\bl{K}',\eta_2\bl{K}-\eta_2\bl{K}'+\bl{k})
 a^\dagger(\bl{K}'-\eta_2\bl{K}-\bl{k})a(\eta_1\bl{K}-\bl{k}) \nonumber \\ & &
 -\int d^3k g(\bl{K},\bl{k})g^*(\bl{K}',\eta_1\bl{K}'-\eta_1\bl{K}+\bl{k})
 b^\dagger(\bl{K}'-\eta_1\bl{K}+\bl{k})b(\eta_2\bl{K}+\bl{k}) \label{cc2} 
\end{eqnarray}
where the normalization of the momentum-space wave function is assumed to be 1. The spectral condition forbids the last two terms on the right-hand side to have matrix elements within the subspace of the bound state.\footnote{
The reason for this is that the operator $b^\dagger b$, for example, in (\ref{cc2}) changes the momentum of the electron for $\bl{K}\neq \bl{K}'$ without changing the momentum of the nucleus, and thus the resulting total momentum cannot satisfy the energy-momentum relation of the bound state. For the exceptional case $\bl{K}= \bl{K}'$, the last two terms in (\ref{cc2}) give a finite correction to $\delta(0)$, which should be ignored.}
 We therefore ignore them and obtain the canonical commutation relations for the asymptotic atom field. 

We have derived the canonical commutation relations for the bound-state fields. This does not, of course, represent a point-like particle picture, because $A^\dagger(\bl{K})$ creates a particle in the plane-wave state. The point-particle picture is relevant only for the interactions.

\section{Anyon fields in the 2 spatial dimensions \label{any}}

The anyon is a particle-like excitation in a 2+1 dimensional system, which is observed, for example, in the phenomenon of the FQHE. It is characterized by fractional statistics in which the interchange of two anyons can result in any phase of the wave function.\cite{Wilczek82} 
The electron system in the FQHE is confined to some 2-dimensional surface by complicated electromagnetic interactions with the surrounding material. The most important effect of the confinement is a change of the topological structure of the configuration space. In the (2+1)-dimensional field theory, this boundary condition is accounted for by including a CS gauge term in the Lagrangian, which is a mathematical device to replace the confining interaction and leads to exotic statistics.\cite{Semenoff88}

\subsection{Exotic statistics in Chern-Simons field theory}

We consider two species of charged particles, whose fields are denoted by $\psi$ and $\phi$. We assume, for definiteness, bosonic commutation relations among them and call them the ``background bosons''. These bosons interact with the CS fields, for which we introduce three CS terms in the Lagrangian, following Ezawa et al.\cite{Ezawa92}\cite{Wilczek92} Then, the CS part of the Lagrangian becomes

\begin{eqnarray}
\mathcal{L}_{\small \mbox{CS}} & = &
    (\dd_\mu+ia_\mu)\psi^*(\dd^\mu-ia^\mu)\psi-m^2\psi^*\psi
  -\f{1}{4\alpha}\varepsilon^{\mu\nu\lambda}a_\mu\dd_\nu a_\lambda \nonumber \\
       &  & +(\dd_\mu+ib_\mu)\phi^*(\dd^\mu-ib^\mu)\phi-M^2\phi^*\phi
  -\f{1}{4\beta}\varepsilon^{\mu\nu\lambda}b_\mu\dd_\nu b_\lambda \nonumber \\
        & & -\f{1}{4\gamma}\varepsilon^{\mu\nu\lambda}(a_\mu\dd_\nu b_\lambda +
                           b_\mu\dd_\nu a_\lambda),     \label{Lcs}
\end{eqnarray}
where the last term governs the mutual statistics between the fields $\psi$ and $\phi$. We quantize this system by following the procedure developed by Semenoff.\cite{Semenoff88}\cite{Matsuyama89} The constraint conditions then become

\begin{eqnarray}
 \f{1}{2\alpha}\varepsilon_{ij}\dd_ia_j+\f{1}{2\gamma}\varepsilon_{ij}\dd_ib_j,
        & = & j_0,                \label{consta} \\
 \f{1}{2\gamma}\varepsilon_{ij}\dd_ia_j+\f{1}{2\beta}\varepsilon_{ij}\dd_ib_j 
        & = & k_0 ,
\end{eqnarray}
where $j_0$ and $k_0$ are the 0th components of the currents of $\psi$ and $\phi$ respectively. Under these conditions, we obtain a Hamiltonian in which $\psi$ and $\phi$ couple minimally to $a_i$ and $b_i$ (i=1,2). Further, by assuming the gauge conditions $\dd_ia_i=\dd_ib_i=0$ we find that the CS fields are given by

\begin{eqnarray}
 a_i(x) & = & \f{1}{\pi}\dd_i\int d^2y\Omega(\bl{x}-\bl{y})
           \{\mu_aj_0(y)-\mu k_0(y)\}\equiv \dd_i\Theta_a(x), \label{ai} \\
 b_i(x) & = & \f{1}{\pi}\dd_i\int d^2y\Omega(\bl{x}-\bl{y})
           \{\mu_bk_0(y)-\mu j_0(y)\}\equiv \dd_i\Theta_b(x), \label{bi} \\
 \Omega(\bl{x}-\bl{y}) & = & \mbox{arctan} \f{x^2-y^2}{x^1-y^1}.
\end{eqnarray}
The coefficients $\mu_a$, $\mu_b$ and $\mu$ are given by

\begin{equation}
 \mu_a=\f{\alpha\gamma^2}{\gamma^2-\alpha\beta},\quad
\mu_b=\f{\beta\gamma^2}{\gamma^2-\alpha\beta},\quad
  \mu=\f{\alpha\beta\gamma}{\gamma^2-\alpha\beta}.    \label{mus}
\end{equation}

We see in (\ref{ai}) and (\ref{bi}) the characteristics of the CS field embodying the boundary condition: We can eliminate the CS gauge interactions from the Hamiltonian by applying the (singular) gauge transformations

\begin{equation}
 \psi(x)=\exp(i\Theta_a(x))\hat{\psi}(x), \quad 
    \phi(x)=\exp(i\Theta_b(x))\hat{\phi}(x) \quad \mbox{etc.} \label{anytra}
\end{equation}
and obtain anyons as follows. By applying (\ref{anytra}) the equal-time commutation relations become exotic

\begin{eqnarray}
\hat{\psi}(x)\hat{\psi}(y)-e^{i(2n+1)\mu_a}\hat{\psi}(y)\hat{\psi}(x) &=&0,
                                                            \nonumber \\
\hat{\psi}(x)\hat{\pi}(y)-e^{i(2n+1)\mu_a}\hat{\pi}(y)\hat{\psi}(x)
                   &=& i\delta(\bl{x}-\bl{y}),  \nonumber \\
\hat{\phi}(x)\hat{\phi}(y)-e^{i(2n'+1)\mu_b}\hat{\phi}(y)\hat{\phi}(x) &=&0,
                                                            \nonumber \\
\hat{\phi}(x)\hat{\chi}(y)-e^{i(2n'+1)\mu_b}\hat{\chi}(y)\hat{\phi}(x)
                   &=& i\delta(\bl{x}-\bl{y}),     \nonumber \\
\hat{\psi}(x)\hat{\phi}(y)-e^{i(2n''+1)\mu}\hat{\phi}(y)\hat{\psi}(x) &=&0,
                                                       \nonumber \\
     & \mbox{etc.} &               \label{psiphi}
\end{eqnarray}
Here, $\pi$ and $\chi$ are the fields canonically conjugate to $\psi$ and $\phi$, respectively. The odd integers $2n+1$, etc., come from the multi-valued nature of the function $\Omega$. Thus, we have the anyon fields $\hat{\psi}(x)$, $\hat{\phi}(x)$, etc. We note that the exclusion principle holds for anyons.

\subsection{Composite anyon field \label{company}}

If we express the free anyon fields as superpositions of the plane waves

\begin{eqnarray}
 \hat{\psi}(x) & = & \int\f{d^2k}{2\pi}\f{1}{2\omega}
           \{a(\bl{k})e^{-ikx}+c^\dagger(\bl{k})e^{ikx}\},  \\
 \hat{\phi}(x) & = & \int\f{d^2k}{2\pi}\f{1}{2\omega}
                 \{b(\bl{k})e^{-ikx}+d^\dagger(\bl{k})e^{ikx}\},
\end{eqnarray}
the commutation relations among the operators $a$, $b$, $c$... become

\begin{eqnarray}
 a(\bl{k})a(\bl{k}')-e^{i(2n+1)\mu_a}a(\bl{k}')a(\bl{k}) &=& 0, \nonumber \\
 a(\bl{k})a^\dagger(\bl{k}')-e^{i(2n+1)\mu_a}a^\dagger(\bl{k}')a(\bl{k})
                            & = & \delta(\bl{k}-\bl{k}'), \nonumber \\
 b(\bl{k})b(\bl{k}')-e^{i(2n'+1)\mu_b}b(\bl{k}')b(\bl{k}) & = & 0,
                                                        \nonumber \\
  b(\bl{k})b^\dagger(\bl{k}')-e^{i(2n'+1)\mu_b}b^\dagger(\bl{k}')b(\bl{k})
                            & = & \delta(\bl{k}-\bl{k}'), \nonumber \\
 a(\bl{k})b(\bl{k}')-e^{i(2n''+1)\mu}b(\bl{k}')a(\bl{k}) &=& 0, \nonumber \\
                   & \mbox{etc.} &  \label{abcom}
\end{eqnarray}

When the values $\mu_a/\pi$, $\mu_b/\pi$ and $\mu/\pi$ are integers, we can construct the Fock space and obtain the particle picture of anyons. We note that the condition under which anyons allow the particle interpretation can be relaxed somewhat, as is discussed in \S\S \ref{anypart} and \ref{FQHE}.

Now, let us assume that two anyon fields with the statistical parameters $\alpha$ and $\beta$ form a composite system. Relying on the ``glue is unimportant'' principle, we can calculate the commutators of the composite operators in the same way as in \S \ref{atom}, where the anticommutators are replaced by their exotic counterparts given in (\ref{abcom}).
 The composite operator $A$ in the space-like asymptotic limit is defined by
 
 \begin{equation}
      A(K)=\int g(K,k)a(\eta_1\bl{K}-\bl{k})b(\eta_2\bl{K}-\bl{k}) d^2k.
 \end{equation}
 With the condisions discussed in \S \ref{atom}, $A$ satisfies the commutation relations

\begin{eqnarray}
  A(\bl{K})A(\bl{K}')-e^{i(2n+1)\mu_{ab}}A(\bl{K}')A(\bl{K}) & = & 0,  \\
   A(\bl{K})A^\dagger(\bl{K}')-e^{i(2n+1)\mu_{ab}}A^\dagger(\bl{K}')A(\bl{K})
    & = & \delta(\bl{K}-\bl{K}'),
\end{eqnarray}
where\cite{endnote}

\begin{equation}
  \mu_{ab}=\mu_a+\mu_b \quad (\mbox{mod}\, 2\pi).      \label{muab}
\end{equation}

\subsection{quasi-particle picture of the anyon    \label{anypart}}
    
We have assumed that the background fields $\psi$ and $\phi$ have one unit of CS (statistical) charge in the Lagrangian (\ref{Lcs}). The CS flux is, therefore, quantized in unit of $2\pi$.
\footnote{We are using natural unit in which $\hbar=1$.} 
Assume an object (anyon) $a$ with CS charge $m_a$ carrying $f_a$ units of flux that is perpendicular to the confined surface. If two such objects exchange their positions by circulaing around each other, the two-particle wave function aquires a phase $\exp(\pm if_am_a\pi)$ by virtue of the Aharonov-Bohm effect for the CS flux. By comparing with (\ref{psiphi}), we are tempted to identify $f_am_a$ with $\mu_a/\pi$ . However, it can be shown that this is not always possible by inspecting the composite anyon state. It is natural to assume the conservation of both the charge and the flux in the process of the binding. Then, the charge and the flux of which $\mu_{ab}$ consists are the sums of those possessed by the anyons $a$ and $b$. But this assignment is not always compatible with (\ref{muab}), as shown in the following.

We first note that anyonic particles cannot coexist if they have different charge/flux ratios, since we cannot define the interchange of two anyons consistently for them. Consider, then, $n$ anyons with the flux $f_i$ and charge $m_i$ that satisfy the relations

\begin{equation}
 m_i=k f_i, \quad i=1,2,...,n
\end{equation}
with a common constant $k$. Suppose next that these $n$ anyons form a composite state, which has flux $\sum_i f_i$ and the charge $\sum_i m_i$. Equation (\ref{muab}) then gives the consistency condition

\begin{equation}
 k\left\{\left( \sum_{i=1}^nf_i\right)^2-\sum_{i=1}^nf_i{}^2\right\}=2p,
                                       \label{sumcon}
\end{equation}
where $p$ is an integer. 
For $n$ identical anyons, this condition becomes $n(n-1)f_1m_1=2p$.

 Though anyon cannot generally be regarded as a particle, we can assign it flux $f_L$ and charge $m_L$ if the converted statistical parameter $\mu_L$ satisfies
 
\begin{equation}
  \f{\mu_L}{\pi}=p=f_Lm_L,
\end{equation}
for an integer $p$, because we can construct the Fock space for this value of the phase. Now, suppose that this anyonic particle decays into $n$ anyons satisfying the criterion (\ref{sumcon}). We can also regard these products of dissociation to be quasi-particles, since their parent state has the particle attributes.

\subsection{Anyons in fractional quantum Hall effects  \label{FQHE}}

The phenomenology of the FQHE becomes simple if we use quasi-particle concepts of anyons, when bosonic background fields are assumed. The ground state (Laughlin state) of the FQHE is an incompressible quantum fluid made of the background boson with charge $-e$ carrying $f_L$ units of the magnetic flux,\cite{Laughlin83} where the quantization unit is $\phi_0=2\pi/e$. We identify it as an anyonic particle with the statistical parameter $\alpha_L$ given by

\begin{equation}
 \f{\alpha_L}{\pi}=f_L=\f{B}{\phi_0\langle\rho\rangle_L},
                                                       \label{alphl}
\end{equation}
where $\langle\rho\rangle_L$ is the boson density and $B$ is the external magnetic field. We see that the filling factor $\nu^L$ is $1/f_L$ in this state. The CS field associated with the boson is given through (\ref{consta}) with $\alpha=\alpha_L$, $\gamma=\infty$ and $j_0=\rho$, which becomes

\begin{equation}
\langle\rho\rangle_L=\f{1}{2\alpha_L}\varepsilon_{ij}\dd_i\langle a_j\rangle_L.
\end{equation}
We therefore have $\langle a_i\rangle_L=eA_i$ for the ground Laughlin state, where $\varepsilon_{ij}\dd_iA_j=B$. Thus, the assumed CS field is determined completely by the real external field in this state.

Suppose that an elementary topological vortex is excited in a magnetic field $B$ increasing in time.\cite{Ezawa92} It is given by

\begin{equation}
 \rho=\langle\rho\rangle_L+ \rho_q, \quad a_i=eA_i+v_i, \quad
                      v_i \to\dd_i\theta \quad (\mbox{at infinity}),
\end{equation}
where $\theta$ is the azimuthal angle in the frame whose origin is at the center of the vortex. The CS flux of this excitation becomes $2\pi$, which amounts to a unit flux $\phi_0$ of the real magnetic field. On the other hand, the statistical charge $Q$ of the field $q$ is given by

\begin{equation}
 Q=\int\rho_qd^2x=\f{1}{2\alpha_L}\int\varepsilon_{ij}\dd_iv_jd^2x
                     =\f{\pi}{\alpha_L}=\f{1}{f_L}.
\end{equation}
Thus, the vortex is identified with a quasi-hole (anti-quasi-particle) having unit flux and CS charge $m_q=1/f_L$. We, further, find that the statistical parameter of the field $q$ is given by $\alpha_q=m_q\pi$, because the anyon $q$ is a dissociation product of another anyon which can be identified with a particle.

The accumulated quasi-holes form a second incompressible-fluid state, where the next vortices are created, which again form a higher Laughlin state, and so on. Though the anyons in this hierarchical structure cannot be regarded as products of the simple dissociation, they are the quasi-particles with definite flux and  CS charge, as is shown in \S \ref{hier}. 

\section{A scalar field in the 1 spatial dimension \label{one}}

In this section, we briefly study statistics in 1 spatial dimension for completeness.

The massless scalar field (Nambu-Goldstone boson) in the Schwinger\cite{Schwinger62} and the Thirring\cite{Thirring58} models was investigated by Ito\cite{KRIto75} and Nakanishi.\cite{Nakanishi75} Ito gave an explicit model of the operators for this composite field. The annihilation operator is given by

\begin{equation}
  d^+(p^1)=\int \f{dq^1}{p^0}\{\theta(p^1) :\Psi_1^*(q^1)\Psi_1(p^1+q^1):
             + \theta(-p^1) :\Psi_2^*(q^1)\Psi_2(p^1+q^1):\}.
\end{equation}
Here, 
\[  \Psi(p^1)=u(p^1)a(p^1) +v(p^1)b^*(-p^1),  \]

\begin{equation}
  u(p^1)=\left(
	\begin{array}{c}\theta(p^1) \\ \theta(-p^1)
\end{array}
\right), \quad v(p^1)=\left(
	\begin{array}{c}\theta(-p^1) \\ \theta(p^1)
\end{array}
\right),                   \label{spn}
\end{equation}
where $\theta(x)$ is the step function and $a$ ($b$) is the annihilation operator for the original fermion (its antiparticle).\footnote{
There is an error in Eq.(5$\cdot$3) of the original paper. We have corrected the definition of $d^+(p^1)$ here. (See Ref. \cite{KRIto75} for further details.)}
 This operator exactly satisfies the canonical commutation relations
\[ [ d^+(p^1), d^+(p'^1)^*]=\delta(p^1-p'^1), \quad \mbox{etc.}  \]
This is due to the special nature of the 1-dimensional `spinor' (\ref{spn}).

\section{Hierarchy of the FQHE \label{hier}}

We found in \S \ref{FQHE} that the CS field is equal to the real field in the Laughlin state. This is an advantage of the assumption that the background field is bosonic, and this mechanism continues to exist in higher parts of the hierarchy. We then have two strategies, depending on the way we treat the CS field. The first one is developed in the next subsection, where we perform the anyonic gauge transformation to eliminate it at every stage of the hierarchy. In the second strategy, briefly described in \S \ref{EHI}, the CS fields and the background bosons are all eliminated succesively by chain relations among them and only physical fields remain in the final results.

\subsection{The hierarchy in terms of the anyonic gauge transformation \label{and}}

Let us begin with the $s$th quasi-particle of the density $\rho_s$, which has charge $q_se$ and statistical parameter $\alpha_s$. The quasi-particles form a Laughlin state with density $\langle \rho\rangle_{s+1}$. Then, the ($s+1$)th vortex is excited, and is given by the configuration

\begin{equation}
 \rho_s=\langle \rho\rangle_{s+1}+ \rho_{s+1}, \quad a_{i(s)} 
                      =\langle a_i\rangle_{s+1} +v_{i(s+1)},   \label{vortex}
\end{equation}
\begin{equation}
         v_{i(s+1)}  \to \tau_{s+1}\dd_i\theta \quad \mbox{(at infinity)},
\end{equation}
where the vortex with $\tau_{s+1}=-1 (1)$ is identified with the quasi-hole (quasi-particle). We assume that the frozen part $\langle\rho\rangle_{s+1}$ satisfies

\begin{equation}
 \langle\rho\rangle_{s+1}=
      \f{1}{2\alpha_s}\varepsilon_{ij}\dd_i\langle a_j\rangle_{s+1}.
\end{equation}
Then, the constraint equation gives

\begin{equation}
 \rho_{s+1}=\f{1}{2\alpha_s}\varepsilon_{ij}\dd_iv_{j(s+1)}.
\end{equation}
Now, the ($s+1$)th CS charge is given by

\begin{equation}
 m_{s+1}=\int \rho_{s+1}d^2x=-\tau_{s+1}\f{\pi}{\alpha_s}.
\end{equation}
The parameter $m_{s}$ determines the statistics, which is also governed by the parameter $\alpha_s$, which is defined by $\mu_a$ in (\ref{mus}) with $\gamma=\infty$. We, therefore, identify $\alpha_s/\pi$ with $m_{s}$ and obtain the reciprocal relation\cite{Ezawa92}

\begin{equation}
 m_{s+1}=-\f{\tau_{s+1}}{m_s}+2p_{s+1},   \label{recm}
\end{equation}
where $p_{s+1}$ is an integer representing the multivalued nature of the statistical factor.

The Coulomb forces act among the quasi-particles of (\ref{vortex}) and the lowest energy state becomes the incompressible fluid. In order to estimate the charge $q_{s+1}$ of the anyons, we next investigate the structure of this fluid state.\cite{Ando93} The $N$ quasi-particles form the ($s+1$)th Laughlin fluid, whose wave function is given by

\begin{equation}
 \Psi_{s+1} \propto \prod_{i<j}(z_i-z_j)^{m_{s+1}}
            \exp\left(-\sum_{i=1}^N\f{|z_i|^2}{4\ell^2}\right),   \label{Laugh}
\end{equation}
where $z_i=x_i+iy_i$, and the index $m_{s+1}$ comes from the anyonic statistics.
 The quantity $1/2\pi\ell^2$ is the state density of the quasi-particle, which is $|q_s|$ times that of the electron. In the wave function (\ref{Laugh}) $M=N m_{s+1}$ is the highest angular momentum of the constituent quasi-particle, and the area occupied by the system is given by $2\pi\ell^2Nm_{s+1}$. Now, the buried quasi-particle is manifested as the excitation in this fluid. The quasi-hole excitation is expressed by the operation $S(z_0)$ on the wave function, since it has one unit of CS flux:

\begin{equation}
 S(z_0)\Psi_{s+1}\propto \prod_i(z_i-z_0)\Psi_{s+1}.
\end{equation}
The area of the system increases by $2\pi\ell^2$ as a result of this operation, which is the $1/m_{s+1}$ of the area occupied by a constituent $s$th quasi-particle. Since the quasi-hole excitation has charge opposite to that of the constituent, we have the charge relation

\begin{equation}
 q_{s+1}=\tau_{s+1}\f{q_s}{m_{s+1}}.    \label{recq}
\end{equation}

Returning to (\ref{Laugh}), we next note that the number of one-quasi-hole states is given by $Nm_{s+1}$ for large $N$. Since the constituent quasi-particle has charge $q_se$, the number of the electron-equivalent states (the number of the flux quanta) becomes $Nm_{s+1}/|q_s|$. On the other hand, since the total charge of the quasi-particles is $-\tau_{s+1}Nq_se$, the charge fraction per electron-equivalent state is $-\tau_{s+1}q_s|q_s|/m_{s+1}$. By considering the change in the charge fraction, we have the recurrence formula for the filling factor $\nu^s$

\begin{equation}
 \nu^{s+1}=\nu^s +\tau_{s+1}\f{q_s|q_s|}{m_{s+1}}.  \label{recf}
\end{equation}

We finally obtain the hierarchy of the FQH states from (\ref{recm}), (\ref{recq}) and (\ref{recf}).\cite{Halperin84}

\subsection{The hierarchy derived from the replacement mechanism \label{EHI}}

We have assumed the anyon transformation (\ref{anytra}) and eliminated the CS fields at every level of the hierarchy in \S \ref{and}. Another way to realize the hierarchy of the filling factors is to work throughout with the background bosons. We have chain relations including the CS fields in this case, and all unphysical fields (the bosons and the CS fields) disappear through successive replacement. Such an approach was taken by Ezawa, et al.\cite{Ezawa92} in the framework of a nonrelativistic model. They obtained a hierarchy in which the filling factors are represented by continued fractions.\cite{Haldane83}

\subsection{The limit of even denominators: neutral Fermi-gas states}

It has been argued that the composite fermion theory\cite{Jain90} predicts the Fermi liquid states with the even-denominator filling factors $\nu =1/2m.$\cite{HLR93} The existence of these states has been regarded as the most powerful evidence for this theory. However, it should be noted that the noninteracting fermions states can also be deduced from the limit of some hierarchy series in the background-boson gauge.

Let us consider the simplest case of $\nu=1/2$. If we set $\tau_s=1$, $p_1=2$ and $p_s=1\ (s>1)$ in the recurrence relations (\ref{recm}), (\ref{recq}) and (\ref{recf}) we obtain

\begin{equation}
  m_s=\f{2s+1}{2s-1}, \quad q_s=\f{1}{2s+1} \quad \mbox{and} \quad
        \nu^s=\f{s}{2s+1}.
\end{equation}
Thus, the $s\to \infty$ limit of the series gives a state of gas consisting of neutal fermions.

The origin of the Fermi gas state differs essentially from that of the Fermi liquid which is predicted by the composite fermion theory. In the latter, it is said that the external magnetic field is cancelled by the CS field, and then the fermions become noninteracting, except for their mutual interactions. Contrastingly, the above anyons become noninteracting because they are neutral. In the composite fermion theories, it would appear that the fact that the CS field is a fictitious field has been overlooked.

\section{Discussion}

We have derived the canonical commutation relations for the 3-dimensional bound-state field. Interpretation of the statistical properties of the bound state from the viewpoint of particle quantum mechanics is not enough persuasive. Instead, we should first conceive of the composite field consisting of the constituent fields as an attribute being acquired by the space-time points. The interpretation of this in the language of particle quantum mechanics may be that the constituents loose their individuality in the bound state and behave as a quantum mechanical unity.\cite{Ito98}

In the hierarchy of the compositeness, the Cooper pair of the superconductor may be in the highest class. The annihilation operator for it is written as

\[ \Phi=\sum_{\blsmall{ k}}A(\bl{k})b_\uparrow(\bl{k})b_\downarrow(-\bl{k}). \]
We then have
\[ [\Phi,\Phi]=0 \quad \mbox{and} \quad [\Phi,\Phi^\dagger]=1  \]
under a restriction similar to that leading to (\ref{cc2}) where the last two terms that represent possible decays of the bound state are ignored.  We have seen that a composite field in 1 spatial dimension satisfies the canonical commutation relations without restricting the Fock space of the original fermion. This is consistent with the stability of the Nambu-Goldstone boson.

The difference between field theory and many-particle quantum mechanics is more prominent in the 2 spatial dimensions. The quantized anyon fields do not generally have a Fock-space representation. Therefore, we cannot conceive of the particle picture of the anyon, except for in very special cases. We may at best define the anyon of quasi-particle as an object having definite flux and charge. However, we do not have a Fock space for this quasi-particle field; that is, the number of quasi-particles is not generally a good quantum number. In such special cases as the FQHS the number of anyons becomes definite.

For a composite state of two anyons, we have obtained the proper anyonic commutation relations. The phase exponents in the commutation relations of the constituent anyons add up to that of the composite anyon, as expected. This relation gives some restrictions for the CS charge and flux of the anyon, which is the decay product of a quasi-particle anyon.

The complex nature of the anyon comes from the CS gauge interaction, which mathematically represents the plane (surface) geometry of the system. The advantage of the background-boson gauge is that the statistical parameters and the CS fields at each stage of the FQHS are determined by the physical parameters and field. Further, the background bosons which have been introduced as fictitious fields are converted into the physical anyons by virtue of the singular gauge transformation. Then, the commutation relations impose the exclusion principle, except for in the special case of the bosonic anyon, on the many-body states of the anyons. The generalized Laughlin states are formed as a consequence of this principle(an analogue of the Fermi degeneracy).

On the other hand, if we use the background-fermion gauge, the CS fields are not completely determined by the physical field in the FQHE, and therefore fictitious fields remain in the final results. This reduces our confidence in interpreting them. Though the CS field cancels the external field for the half-filling state, it remains fictitious. The real reason for the disappearance of the interaction cannot be this cancellation.

\section*{Acknowledgements}

The author would like to express his thanks to Professor Nakanishi for calling his attention to the 1-dimensional case.
 He was stimulated by discussions on BEC in the workshop ``Thermo Field Dynamics" held at RIFP in the summers of 1999 and 2000. He would like to thank the organizers and participants with whom he enjoyed having discussions.



\begin{thebibliography}{99}
\bibitem{Zimmermann58}
W. Zimmermann.
\newblock {    Nuovo Cim. {\bf 10}(1958), 597}.
\bibitem{Sakata61}
S. Sakata.
\newblock {    Prog. Theor. Phys. Suppl., {\rm No 19}(1961), 3}.
\bibitem{Ehrenfest30}
P. Ehrenfest and J.R. Oppenheimer.
\newblock {    Phys. Rev. {\bf 37}(1931), 333}.
\newblock {The author is indebted to Prof. O.W. Greenberg for mailing information about his paper which refers this classical paper.}
\bibitem{Haag58}
R. Haag.
\newblock {Phys. Rev. {\bf 112}(1958), 669}.
\bibitem{Ito82}
H. Ito.
\newblock{Prog. Theor. Phys.{\bf 67}(1982), 1553};
\newblock { ibid.{\bf 78}(1987), 978};
\newblock { ibid.{\bf 89}(1993), 763}.
\bibitem{Wilczek82}
F. Wilczek.
\newblock {    Phys. Rev. Lett. {\bf 49}(1982), 957}.
\bibitem{Semenoff88}
G. W. Semenoff.
\newblock {    Phys. Rev. Lett. {\bf 61}(1988), 517}. 
\bibitem{Ezawa92}
Z. F. Ezawa, M. Hotta and A. Iwazaki.
\newblock {    Phys. Rev. {\bf B46}(1992), 7765}. 
\bibitem{Wilczek92}
F. Wilczek.
\newblock {    Phys. Rev. Lett. {\bf 69}(1992), 132}. 
\bibitem{Matsuyama89}
T. Matsuyama.
\newblock {    Phys. Lett. {\bf B228}(1989), 99}. 
\bibitem{endnote}
\newblock{We have at first $\mu_{ab}=\{(2n+1)\mu_a+(2n'+1)\mu_b\}/(2\bar{n}+1)$. We should put $\bar{n}=n=n'$ since the left hand side is independent of these integers. Similar considerations of consistency are necessary in treating the multi-valuedness in (\ref{psiphi}).}
\bibitem{Laughlin83}
R. B. Laughlin.
\newblock {    Phys. Rev. Lett. {\bf 50}(1983), 1395}. 
\newblock{We simply consider the Laughlin state as the product of an analogue of the Fermi degeneracy rather than the Bose condensation.}
\bibitem{Schwinger62}
J. Schwinger.
\newblock {    Phys. Rev. {\bf 128}(1962), 2425}.
\bibitem{Thirring58}
W. E. Thirring.
\newblock {    Ann. of Phys. {\bf 3}(1958), 91}.
\bibitem{KRIto75}
K. R. Ito.
\newblock {    Prog. Theor. Phys. {\bf 53}(1975), 817}.
\bibitem{Nakanishi75}
N. Nakanishi.
\newblock {    Prog. Theor. Phys. {\bf 54}(1975), 840} ;
\newblock {    ibid. {\bf 57}(1977), 580}.
\bibitem{Ando93}
T. Ando.
\newblock {Iwanami series of {\it The Current Physics {\bf 18}}(in Japanese) (Iwanami Shoten, Tokyo, 1993), p.137. }

\bibitem{Halperin84}
B. I. Halperin.
\newblock {    Phys. Rev. Lett. {\bf 52}(1984), 1583, ibid. {\bf 52}(1984), 2390}.
\newblock{The initial conditions are $q_0=m_0=\tau_1=1$ and $\nu^0=0$.}
\bibitem{Haldane83}
F. D. M. Haldane.
\newblock {    Phys. Rev. Lett. {\bf 51}(1983), 605}.
\bibitem{Jain90}
J.K. Jain.
\newblock {    Phys. Rev. {\bf B41}(1990), 7653}.
 For reviews,

S.D. Sarma and A. Pinczuk(eds.).
\newblock{{\it Perspectives in Quantum Hall Effects}(John Wiley and Sons, New York, 1997), p.225(B.I. Halperin), p.265(J.K. Jain)}.

Z.F. Ezawa.
\newblock{{\it Quantum Hall Effects}( World Scientific, Singapore, 2000), p.268.}
\bibitem{HLR93}
B.I. Halperin, P.A. Lee and N. Read.
\newblock {    Phys. Rev. {\bf B47}(1993), 7312}.
\bibitem{Ito98}
\newblock{H. Ito.}
\newblock{    J. Fac. Sci. Engg. Kinki Univ. {\bf 34}(1998), 9}.

\end{thebibliography}
\end{document}